\documentclass{aastex}
\usepackage{emulateapj5}

\begin{document}
\title{On the vertical equilibrium of the local Galactic disk and
the search for disk dark matter}
\author{F.~J.~S\'anchez-Salcedo$^{1}$, Chris Flynn$^{2,3}$ and 
A.~M.~Hidalgo-G\'amez$^{4}$} 
\affil{$^{1}$Instituto de Astronom\'{\i}a, Universidad Nacional
Aut\'onoma de M\'exico, Ciudad Universitaria,
Apt.~Postal 70 264, C.P.~04510, Mexico City, Mexico, 
Email:jsanchez@astroscu.unam.mx}
\affil{$^{2}$Department of Physics and Astronomy, University of Sydney, NSW 2006
Australia}
\affil{$^{3}$Finnish Centre for Astronomy with ESO, University of Turku,
FI-21500 Piikkio, Finland}
\affil{$^{4}$Departamento de F\'{\i}sica, Escuela Superior de F\'{\i}sica 
y Matem\'aticas, Instituto Polit\'ecnico Nacional, 
\\ U.P. Adolfo L\'opez Mateos, 
C.P. 07738, Mexico City, Mexico}
\begin{abstract}

Estimates of the dynamical surface mass density at the solar Galactocentric
distance are commonly derived assuming that the disk is in vertical equilibrium
with the Galactic potential. This assumption has recently been called into
question, based on the claim that the ratio between the kinetic and the
gravitational energy in such solutions is a factor of $3$ larger than required
if Virial equilibrium is to hold. Here we show that this ratio between energies
was overerestimated and that the disk solutions are likely to be in Virial
equilibrium after all. We additionally demonstrate, using one-dimensional
numerical simulations, that the disks are indeed in equilibrium.  Hence, given
the uncertainties, we find no reason to cast doubt on the steady-state
solutions which are traditionally used to measure the matter density of the
disk.

\end{abstract}


\section{Introduction}
Studies on the amount of unseen mass associated with the Galactic disk, have
been carried out since Oort (1932) and up to the present day (e.g., Kuijken \&
Gilmore 1991; Bahcall et al.~1992; Cr\'ez\'e et al.~1998; Siebert et al.~2003;
Holmberg \& Flynn 2004; Kalberla et al.~2007; Moni Bidin et al.~2010).  All
recent studies indicate that there is no dynamically significant amount of disk
dark matter at the solar position, although a small fraction ($<$10\%) cannot
be ruled out.  In order to derive the dynamical mass surface density, the usual
way to proceed is to consider a tracer of the potential, such as K giant stars
or H\,{\sc i} gas.  In principle, ideal tracers are old enough (age $>1$ Gyr)
that they are expected to be well mixed into the Galactic potential, so that
they satisfy the Virial condition. In a recent paper, Garrido Pesta\~na \&
Eckhardt (2010) (hereafter GPE) compare the kinetic energy $T$ of stellar and
gaseous components in the local Galactic disk, to their gravitational energy
$W$, and conclude that the vertical configuration is not in a state of
equilibrium because $|2T/W|\simeq 3$ and, therefore, the classical
Poisson-Boltzmann approach used to interpret the data is an inappropriate and
futile exercise.  If the energy in vertical random motions is a factor of $3$
larger than in Virial equilibrium, the local portion of the disk should be in a
phase of violent relaxation. Under such conditions, the thickness of the disk
would inflate by a factor of $5/3$ on average (see section \ref{sec:steady})
after just a few vertical crossing times ($\sim 10^{8}$ years). It is unlikely
that we would be observing the disk at a time just prior to such a violent
event, and if we were, the classical view that the disk thickening is due to
slow dynamical heating due to encounters with massive molecular clouds, spiral
arms and gravitational perturbations caused by minor mergers, should change
dramatically. In this paper, we show, however, that there is no reason for
abandoning the assumption that the vertical configuration is in a steady state.

\section{Steady-state configuration}
\label{sec:steady}
Consider a rotationally symmetric disk stratified in the
$z$-direction, where $z=0$ is the disk plane of symmetry,
 and composed of $N$ components in equilibrium.
Each component, either if it is collisionless (stars
and dark matter) or collisional (gas), satisfies a Boltzmann equation.
The vertical Jeans equations can be combined to obtain:
\begin{equation}
\sum_{i=1}^{N}  \frac{\partial(\rho_{i}\sigma_{i}^{2})}{\partial z}=
-\rho\frac{\partial \Phi}{\partial z},
\label{eq:1stboltzmann}
\end{equation}
where $\sigma_{i}(z)$ is the vertical velocity dispersion of the component
$i$, $\Phi(z)$ is the gravitational potential and
$\rho(z)\equiv \sum\rho_{i}(z)$. 
For gaseous components, $\sigma_{i}$ is the effective velocity dispersion,
which includes the pressure support by turbulent motions, magnetic fields
and cosmic rays.
In a real galaxy, gradients in the radial and azimuthal
directions generate nondiagonal terms of the velocity dispersion
tensor, i.e.~$\sigma_{Rz}$ and $\sigma_{\theta z}$, in the
vertical Jeans equation. At $z\lesssim 1$ kpc, the fractional
corrections due to the cross terms are less than $1\%$ 
and are usually ignored except when studying stars at $z>2$ kpc
(e.g., Bahcall 1984; Olling 1995; Becquaert \& Combes 1997; Kalberla 2003; 
Moni Bidin et al.~2010).

Following GPE, we start by considering a infinite 
plane-parallel layer of total column density $\Sigma_{\infty}$ 
in equilibrium with its own gravity.
By choosing the zero of the potential at $z=0$ and
multiplying the above equation by $z$ and integrating over $z$,
one obtains the Virial theorem, linking the kinetic and
gravitational energies with surface terms. If the surface terms are
null at $z\rightarrow \pm \infty$, the relationship is $2T=W$, where
$W$ and $T$ are the gravitational and kinetic energies (due to
vertical motions), respectively, per surface unit in the disk\footnote{Note that
$W>0$ in this case. GPE uses a different sign convention.}.
GPE show that the Virial theorem can
be rewritten as $\tilde{R}=Q$ where 
\begin{equation}
\tilde{R}=\frac{\rho(0)\left<\sigma^{2}\right>}{\pi G \Sigma_{\infty}^{2}},
\label{eq:garpes}
\end{equation}
and 
\begin{equation}
Q= \int_{0}^{\infty}\int_{-\infty}^{\infty} ap(z)p(z+a) \,dz\, da,
\end{equation}
where $p(z)=\rho(z)/\Sigma_{\infty}$.
Here $\left<\sigma^{2}\right>$ is the mass-weighted
mean-square velocity dispersion, 
i.e.~$\left<\sigma^{2}\right>\equiv 2T/\Sigma_{\infty}$.

For the particular case of having
a single self-gravitating isothermal component ($N=1$), it holds that $Q$ is
exactly $1/2$,
whereas it is slightly different from $1/2$ for a Gaussian density profile in
equilibrium ($Q=0.45$). GPE argue that $Q\simeq 1/2$
and evaluate the left-hand-side of Eq.~(\ref{eq:garpes})
for the Galactic disk at the solar position
and find that it has the value $1.6$. Since this is a factor of $\sim 3$ larger
than required if the Virial equilibrium is to hold, they 
conclude that the disk at the solar position is not in a 
steady-state vertical configuration\footnote{If true, the disk will 
reach a new (final) equilibrium configuration,
with kinetic energy $T_{f}$ and potential energy $W_{f}$,
such that $E=(3/2)W_{f}$, where $E$ is the kinetic plus
gravitational energies. In the initial nonequilibrium configuration,
$E=T_{i}+W_{i}=(5/2)W_{i}$. Equating the energy, we find that 
$W_{f}=(5/3)W_{i}$.  Since the scaleheight scales roughly as $W$,
the disk will become thicker by a factor of $5/3$.}.

In the following, we will check the robustness of the assumption 
that $Q\simeq 1/2$. 
To do so, we start by integrating Equation (\ref{eq:1stboltzmann}) from
the midplane to a certain distance $z$ above the midplane and obtain:
\begin{equation}
\sum_{i=1}^{N} \rho_{i}(z)\sigma_{i}^{2}(z)=
\sum_{i=1}^{N} \rho_{i}(0)\sigma_{i}^{2}(0)-
\int_{0}^{z} \rho(z') \frac{\partial \Phi}{\partial z'} dz'.
\label{eq:pressure}
\end{equation}
For a system such that the volume density of kinetic energy decays
to zero at large $z$, Eq.~(\ref{eq:pressure}) becomes
\begin{equation}
\sum_{i=1}^{N} \rho_{i}(0)\sigma_{i}^{2}(0)=
\int_{0}^{\infty} \rho(z') \frac{\partial \Phi}{\partial z'} dz',
\label{eq:pressure_midplane}
\end{equation}
and tell us that the total pressure at the
midplane must be in balance with the weight of all the material above $z=0$.

In the particular case of a stratified plane-parallel layer,
all the variables depend only on $z$.
Integration of the Poisson equation allows us to find the vertical
acceleration. If the layer is confined by its own gravity,
the vertical acceleration for $z>0$ reads 
\begin{equation}
\frac{d\Phi}{dz}=2\pi G\Sigma(z),
\label{eq:verticalforce}
\end{equation}
where $\Sigma(z)$ is defined for $z>0$ as
\begin{equation}
\Sigma(z)=\int_{-z}^{z} \rho(z') dz'. 
\end{equation}
Combining Eqs.~(\ref{eq:pressure_midplane}) and (\ref{eq:verticalforce}), and
using the relation
\begin{equation}
\rho(z)=\frac{1}{2} \frac{d\Sigma}{dz},
\end{equation}
we find
\begin{equation}
\sum_{i=1}^{N} \rho_{i}(0)\sigma_{i}^{2}(0)=
\pi G \int_{0}^{\infty} \Sigma(z') \frac{d\Sigma}{dz'} dz'
=\frac{1}{2}\pi G\Sigma^{2}_{\infty},
\label{eq:middle}
\end{equation}
which can be expressed as
\begin{equation}
\frac{\sum_{i} \rho_{i}(0)\sigma_{i}^{2}(0)}{\pi G\Sigma^{2}_{\infty}}=\frac{1}{2}.
\label{eq:exact}
\end{equation}
Note that the above equation is {\it exact} for a plane-parallel layer
in equilibrium.
Therefore, the total pressure at $z=0$ (not 
$\rho(0)\left<\sigma^{2}\right>/(\pi G \Sigma_{\infty}^{2})$ as stated in GPE)
divided by $\pi G\Sigma^{2}_{\infty}$ 
must be $1/2$ in a steady-state plane-parallel layer.
GPE analysis applied to a plane-parallel layer is strictly correct 
only if $\rho(0) \left<\sigma^{2}\right>= \sum \rho_{i}(0)\sigma_{i}^{2}(0)$.
We will see later that this equality does not hold in a multicomponent
disk.

In a real galaxy, not only the disk but also other components (such as
the dark matter halo) contribute to the vertical confinement of stars 
in the thin and thick disks.
From the Poisson equation, the gravitational vertical acceleration
is given by 
\begin{equation}
\frac{\partial \Phi}{\partial z}= 2\pi G [\Sigma_{d} (z)+ \Sigma_{h}(z)]
-2\int_{0}^{z} \Omega (\Omega + G) dz',
\end{equation} 
where $\Sigma_{d}(z)$ and $\Sigma_{h}(z)$ are the column density 
of the disk and of a spherical dark halo, respectively,  
$\Omega^{2}=R^{-1}\partial \Phi/\partial R$ and 
$G\equiv R(\partial\Omega/\partial R)$ (e.g., Kuijken \& Gilmore 1989;
Ferri\`ere 1998). At $R>5$ kpc and within a few kpc from the
disk plane, $\Omega+G\simeq 0$ (e.g., Kuijken \& Gilmore 1989)
and thus
\begin{equation}
\frac{\partial \Phi}{\partial z}=2\pi G \Sigma (z)+ \frac{v_{h}^{2}}{R}
\frac{z}{\sqrt{R^{2}+z^{2}}},
\label{eq:kg89}
\end{equation} 
where $v_{h}(R)$ is the contribution of the dark halo to the rotation curve;
$v_{h}\simeq 140$ km s$^{-1}$ at the solar position (e.g., Flynn et al.~1996; 
Klypin et al.~2002). In the particular case of a massless disk embbeded
in the spherical potential of the isothermal dark halo, we recover the
classical result that the vertical epicyclic 
frequency coincides with the orbital
frequency at the midplane, $\partial^{2}\Phi/\partial z^{2}|_{z=0}=\Omega^{2}$.
Substituting Eq.~(\ref{eq:kg89}) into Eq.~(\ref{eq:pressure_midplane}) 
and for $z\ll R$, we obtain
\begin{equation}
\frac{\sum \rho_{i}(0)\sigma_{i}^{2}(0)}{\pi G\Sigma_{\infty}^{2}}\simeq
\frac{1}{2}+(\ln 2) \left(\frac{v_{h}^{2}/R}{2\pi G\Sigma_{\infty}}\right)
\sum_{i=1}^{N} \left(\frac{h_{i}}{R}\right)\left(\frac{\Sigma_{i}}
{\Sigma_{\infty}}\right),
\label{eq:exact_eq}
\end{equation}
where the index $i$ refers to the disk components, 
$h_{i}$ is the vertical scaleheight of population $i$ and
$\Sigma_{\infty}=\Sigma_{d}(z=\infty)$.

Evaluation of $\sum \rho_{i}(0)\sigma_{i}^{2}$ (not including the stellar halo)
in the updated disk model of Holmberg \& Flynn (2004) reported in Table 2 of
Flynn et al.~(2006) gives $20.2 M_{\odot}$pc$^{-3}$ km$^{2}$ s$^{-2}$.  Taking
the total surface density in disk components of $\Sigma_{\infty}=48.7
M_{\odot}$pc$^{-2}$, the left-hand-side of Equation (\ref{eq:exact_eq}) is
$0.61$. We then compute the right-hand-side of Eq.~(\ref{eq:exact_eq}) at
$R=R_{\odot}=8$ kpc and find that it is $\simeq 0.56$, which is only $9\%$
smaller than the left-hand-side.  
This simple estimate indicates that the disk model of
Flynn et al.~(2006) is very close to satisfying the Virial theorem.  In order
to assess if this difference of ten percent is significant, we have calculated
$\rho_{i}(z)$ in the model of Flynn et al.~(2006) and computed the 
right-hand-side of Eq.~(\ref{eq:exact_eq}) exactly.  
We confirmed that the disk model fulfills the Virial condition.

Why did GPE obtain that the disk models by Bahcall et al.~(1992) and
Flynn et al.~(2006) do not comply with the Virial theorem?
It was a consequence of the combination of three factors:
(1) they adopted $Q=1/2$, which is in error by a factor of $2$ (see below),
(2) the stellar halo was treated as a disk component, which introduces an
additional artificial factor of $1.25$ and (3) the contribution of the
dark halo in the vertical confinement of disk stars was not taken
into account.
As a consequence, the ratio between kinetic and gravitational energies was
overestimated by a factor of $3$.

For a stratified plane-parallel self-gravitating layer, 
it is true that $\tilde{R}=Q$ in Virial
equilibrium. However, $Q$ is not necessarily $1/2$ in a generic disk.
In fact, the form of the left-hand-side
of the equilibrium condition used in GPE 
(Eq.~\ref{eq:garpes}) can be preserved by introducing a
``correction'' factor $\epsilon$ on the right-hand-side:  
\begin{equation}
\frac{\rho(0)\left<\sigma^{2}\right>}{\pi G \Sigma_{\infty}^{2}}=
\frac{\epsilon}{2},
\end{equation}
with
\begin{equation}
\epsilon=\frac{\rho(0)\left<\sigma^{2}\right>}
{\sum \rho_{i}(0)\sigma_{i}^{2}}.
\end{equation}
The value of $\epsilon$ depends on the vertical structure of the disk.
For a single Gaussian (non-isothermal) density profile in $z$,
$\epsilon=\left<\sigma^{2}\right>/\sigma^{2}(0)=2\sqrt{2}/\pi=0.9$, and hence
$Q=0.45$, as reported in GPE.
However, for the $14$th-components of the
updated disk model of Holmberg \& Flynn (2004), 
the $\epsilon$-value in
the Virial theorem criterion for a state of equilibrium is $2$ and thus 
$Q\simeq 1$. 
To be more precise, we adopted the values of $\rho_{i}(0)$ and
$\sigma_{i}$ for the $14$th-components of Flynn et al.~(2006)
and solved the Poisson-Boltzmann equation for
such a disk with no dark halo, in order to compare with GPE analysis
on a common ground. We obtain 
$\rho(0)=0.091 M_{\odot}$pc$^{-3}$, $\Sigma_{\infty}=55 M_{\odot}$pc$^{-2}$,
$\left<\sigma^{2}\right>^{0.5}=22.25$ km s$^{-1}$ and $\tilde{R}=Q=1.1$.
{\it Therefore, GPE analysis applied to the infinite plane-parallel
multicomponent layer overestimates the $T/W$ ratio by a factor of $2.2$ because}
\begin{equation}
\frac{\rho(0) \left<\sigma^{2}\right>}{\sum\rho_{i}(0) \sigma_{i}^{2}}=2.2.
\end{equation}

\section{Simulations}
In the last section we have shown that the Virial theorem is satisfied for the
Galactic disk models under consideration. However, the Virial theorem is a
relation between global quantities and is not a guarantee that the system is
in a steady-state equilibrium. In order to dispel any doubt, we have tested
the temporal stability of the type of disks considered here using N-body
simulations, in which stars are initially set up in a one-dimensional density
and velocity distribution, based on solutions to the Poisson-Boltzmann
equation, after which we calculate their vertical motions under the
gravitational force generated by the overall density distribution of the disk.

At each time step in the simulation, the downward acceleration at height $z$
above the disk, is found by integrating numerically the density profile of all
the stars (Eq.~\ref{eq:verticalforce}) to absolute height $|z|$. A leap-frog
integrator is then used to compute new positions and velocities for the
particles, and the process repeated. We keep track of the velocity dispersion,
scale height and other quantities for the ensemble, to check if it is in
vertical equilibrium or not.

Three disk types were considered. Firstly, we looked at an ``isothermal disk'',
in which the initial conditions are a sech$^2(z/h)$ distribution in density
(with a scale height $h$) and a (constant) velocity dispersion $\sigma(z)$,
given by by $\sigma= h \sqrt{2 \pi G \rho(0)}$, where $\rho(0)$ is the disk
central density. The density distribution in such a disk is given by $\rho(z) =
\rho(0)$sech$^2(z/2h)$. We adopt $\rho(0) = 0.1$ M$_\odot$ pc$^{-3}$ and $h$ =
$250$ pc, implying a total disk surface density of 50 M$_\odot$ pc$^{-2}$ and a
velocity dispersion of $\sigma= 13.0$ km s$^{-1}$. These are similar to the
measured values for the local Galactic disk.

Secondly, we looked at an ``exponential disk'', in which the density
distribution is given by $\rho(z) = \rho(0)$exp$(-|z|/h$), and the velocity
dispersion is a function of $z$, being smaller close to the midplane and
rising asymptotically to a constant value at large height. The velocity
dispersion $\sigma(z)$ is given by
\begin{equation}
\sigma(z) =  2h \sqrt{\pi G \rho(0) (1-0.5\exp(-|z|/h))}.
\end{equation}
Adopting a scaleheight of $h = 250$ pc and a central density of 
$0.1 M_\odot$ pc$^{-3}$, this disk has a surface density of 50 $M_\odot$ pc$^{-2}$,
like the isothermal disk. The central velocity dispersion for this disk rises
from $13.0$ km s$^{-1}$ at $z=0$ to $18.4$ km s$^{-1}$ at large $z$.

The simulations of these two disks were run with 10 million particles, for $5$
Gyr, over 1 Myr time steps. This time step was found to be a good compromise
between the length of the runs and the accuracy of the orbits. Note that the
oscillation time for stars in the tested potentials is typically $30$ to $80$ 
Myr.  Both disks are completely self-gravitating, with the gravitational field
provided by the density distribution of the particles alone.

Over $5$ Gyr, the central density, mean velocity dispersion and scaleheights of
the simulated disks remain within a few percent of their initial values. Thus,
the exponential disk and the isothermal disk are found to be stationary in the
simulations. Furthermore, this is what we expect, since GPE's analysis shows
that, for both simulated disks, the Virial equilibrium condition is met very
precisely. Thus, no long term secular evolution is expected for either disk, and
this we fully confirm with the simulations.

We then simulated the much more complex system, consisting of all components,
stellar and gaseous, of the disk in Flynn et al.~(2006). This disk
consists of 14 isothermal components, with a range of central densities and
(isothermal) velocity distributions. The stellar halo, which is the 15th
component of the Flynn et al.~(2006) disk model, is ignored (including it in the
solutions is found to have no effect on the disk stability, but we focus in the
following on disk components only). The gaseous components are modelled as
collisionless particles, just like the stars, since the aim is to test the
Poisson-Boltzmann solution, which implicitly makes this assumption. We also
include the effect of the halo dark matter in these simulations, not by
simulating it with particles, but by adding it in to the computation of the
vertical force as an extra density term (as a function of $z$).

\begin{figure} %
\epsscale{0.7}
\plotone{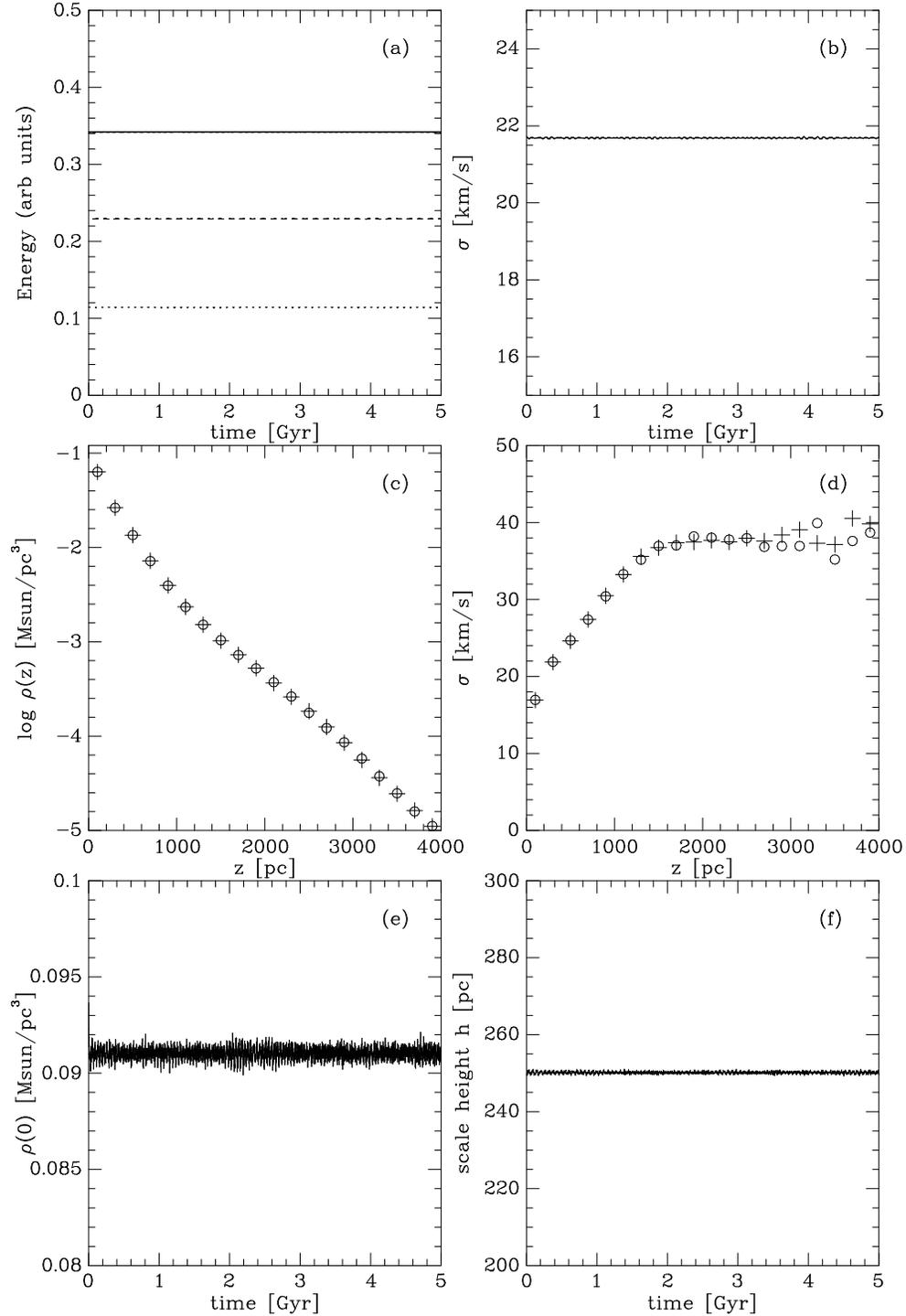}
\caption{Simulation of the secular evolution of the Flynn et al.~(2006) model
  disk. Panel (a) shows the total energy (solid line), the kinetic energy
  (dotted) and the potential energy (dashed) as functions of time, from 0 to 5
  Gyr. Panel (b) shows the total velocity dispersion of all the particles as a
  function of time.  The central panels, (c) and (d), show the initial
  (circles) and final (crosses) density and velocity dispersion distributions
  of the particles as functions of height above the disk $z$; the distributions
  are close to identical, indicating that the disk solutions are stationary and
  that there is no long term evolution in its structure. Panels (e) and (f)
  show the evolution of central density and scale height as functions of
  time. All indicators demonstrate that the disk is in a stationary state in
  the long term.}
\label{fig:hfhalo}
\end{figure}

The simulation is run with 10 million particles for a period of 5 Gyr with time
steps of 1 Myr. Energy conservation in the simulation is better than $0.1$
percent. We find that this complex disk is also stationary. The density and
velocity distributions of the particles after 5 Gyr are very similar to their
initial distributions, and so no large scale secular evolution takes
place. If this rather complex disk were substantially out of equilibrium
initially, we would expect rapid evolution in these quantities. Figure 
\ref{fig:hfhalo} shows
the results. Panel (a) shows the evolution of kinetic energy (dotted line),
potential energy (dashed line) and total energy (solid line) over the 5 Gyr of
the simulation. Panel (b) shows the total velocity dispersion as a function of
time. In panels (c) and (d), we show the initial (circles) and final (crosses)
density and velocity distributions of the particles as a function of $z$
height. The initial and final distributions are essentially indistinguishable.
Panel (e) shows the time evolution of the central density, and panel (f) the
scale height ($h$, of an exponential fit to the density profile, measured from
the midplane to $z \approx 750$ pc). None of the plots show any significant
evolution in these quantities with time. We conclude from the simulations that
the Flynn et al.~(2006) model disk is indeed in equilibrium, as we concluded
from theoretical analysis in the previous section, and contrary to the claims
in GPE.

\begin{figure} %
\plotone{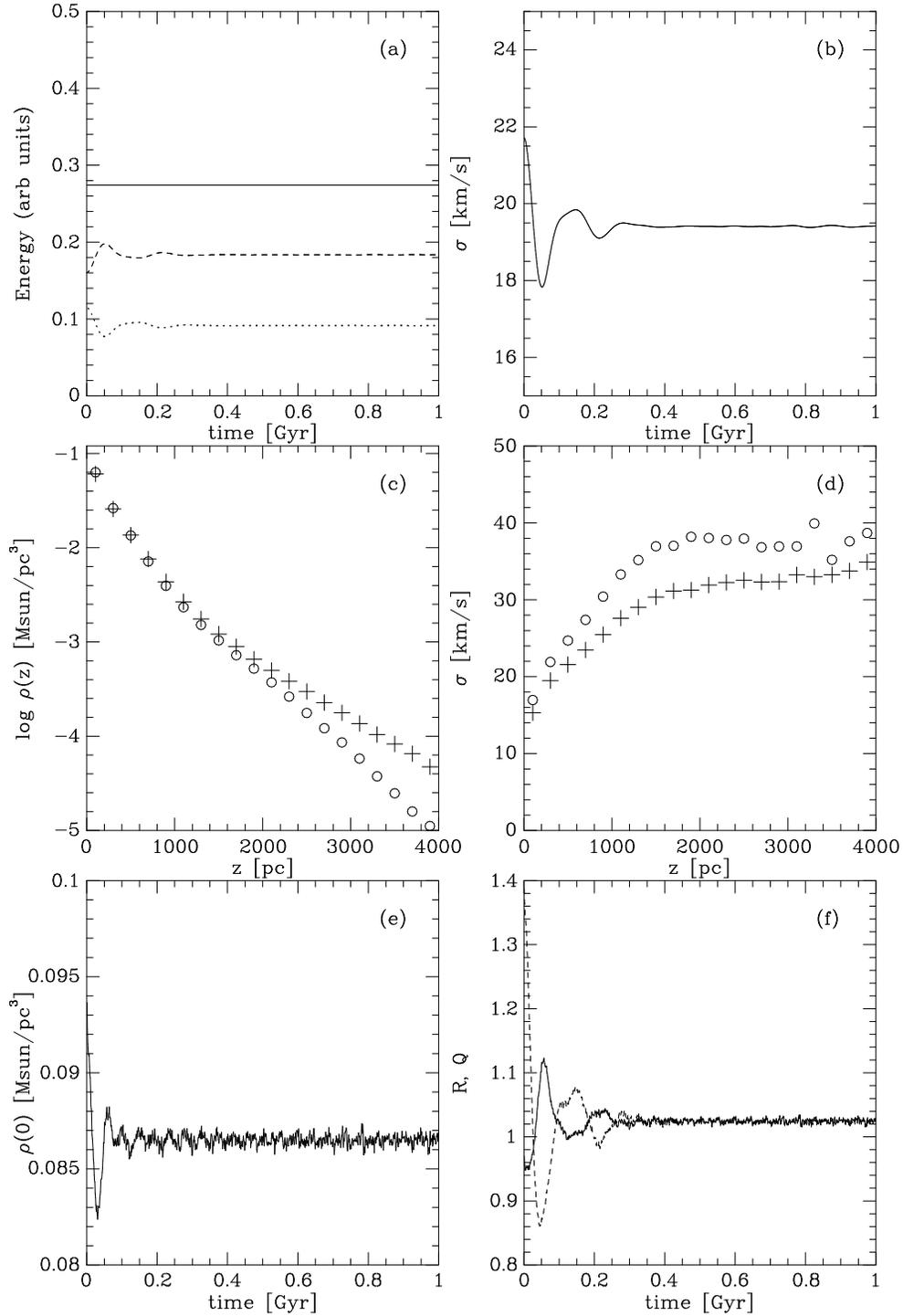}
\caption{Simulation of the secular evolution of the Flynn et al.~(2006) model
  disk, as in Figure \ref{fig:hfhalo}, but with the special case that the dark matter halo has
  been removed from the simulation. All panels are as for 
  Figure \ref{fig:hfhalo}, except
  panel (f), which shows the quantities $\tilde{R}$ (dashed line) and $Q$ (solid line).
  Note that the time axis shows 0 to 1 Gyr, to highlight the rapid response of
  the disk. The disk is initially out of Virial equilibrium, but adjusts itself
  to equilibrium within a few dynamical times, arriving at the state
  $\tilde{R}=Q$. Since the gravity of the dark halo is lacking, which would help to
  confine the disk, the response is for the potential energy to increase which
  kinetic energy decreases, i.e. the disk thickens (panel c) and becomes
  kinematically cooler (panel d).}
\label{fig:unbalanced}
\end{figure}

In order to illustrate the role of the dark halo in confining the disk, we
rerun the simulation but the dark halo contribution to the force was turned off
so that the disk is not in equilibrium initially.  Figure 
\ref{fig:unbalanced} shows that the disk
relaxes dynamically (particularly those stars in the thick disk) to a
stationary state in $~0.3$ Gyr. Panels (a) and (b) show the typical
behaviour of $T$, $W$ and $\left<\sigma^{2}\right>$ in a relaxing process. 
As can be seen from panels (c), (d) and (e),
the disk becomes thicker and colder.
Panel (f) shows the temporal evolution of $\tilde{R}$ and $Q$. 
Since all the gravity is provided by the stars and the gas only, 
Virial theorem for this system dictates that $\tilde{R}=Q$. 
We see that $\tilde{R}=Q=1$ in the steady-state.

\section{Discussion and conclusions}

We have examined the claim by GPE that solutions for the density distribution
of matter in the local Galactic disk are seriously in error because such
solutions are not in Virial equilibrium. GPE claim that the ratio between the
kinetic and the gravitational energy in such disks is a factor of $3$ larger
than required for Virial equilibrium to hold. We have shown that this ratio
between energies was overerestimated by GPE 
because (1) they assumed that $Q=1/2$,
(2) the stellar halo was treated as a disk component
and (3) the vertical confinement provided by the dark halo was ignored.

Taking into account these corrections, the local disk model of Flynn et 
al.~(2006) is likely to be in Virial equlibrium after all. Our technique could
easily be applied to any such model (e.g., Bahcall et al.~1992; Just \&
Jahreiss 2010) and yield a similar result. Simulations of the oscillation of
stars in model disks show that the initial distribution of the stars in density
and velocity space are stationary, backing up our theoretical analysis.

\acknowledgments 
We would like to thank J.~L.~Garrido and the referee for very 
enlightening comments.
CF is very grateful to the University of Sydney, for hosting a visit
where this work was carried out.
This work has been partly supported by CONACyT project CB-60526F.

\clearpage

\clearpage

\end{document}